\documentclass[submitting]{nst}
\usepackage{subfigure,dcolumn}
\usepackage{epstopdf}
\usepackage{mhchem}
\usepackage{upgreek}
\usepackage{amsmath}
\usepackage{xcolor}

\begin{document}
\title{A Two-dimensional Bayesian Approach to Centrality Determination in Nucleus-Nucleus Collisions}
\author{D.\,Idrisov}
\email[Corresponding author, ]{idrisov.dim@mail.ru}
\affiliation{Institute for Nuclear Research of the Russian Academy of Sciences, \\117312, Prospekt 60-letiya Oktyabrya 7a, Moscow, Russia}

\author{F.\,Guber}
\affiliation{Institute for Nuclear Research of the Russian Academy of Sciences, \\117312, Prospekt 60-letiya Oktyabrya 7a, Moscow, Russia}

\author{N.\,Karpushkin}
\affiliation{Institute for Nuclear Research of the Russian Academy of Sciences, \\117312, Prospekt 60-letiya Oktyabrya 7a, Moscow, Russia}

\author{P.\,Parfenov}
\affiliation{Institute for Nuclear Research of the Russian Academy of Sciences, \\117312, Prospekt 60-letiya Oktyabrya 7a, Moscow, Russia}
\affiliation{National Research Nuclear University MEPhI, \\115409, Kashirskoe sh. 31, Moscow, Russia}
\affiliation{Joint Institute for Nuclear Research, \\141980, Joliot-Curie St. 6, Dubna, Russia}

\begin{abstract}
The determination of centrality in nucleus-nucleus collisions is a crucial task, 
as it enables the estimation of the impact parameter and thereby allows for the comparison of experimental results with predictions from theoretical models and other experiments. In this work, we present a two-dimensional approach for centrality determination based on a Bayesian framework. The observables used 
were the number of track hits and the deposited energy of spectators in the 
forward hadronic calorimeter. A distribution is proposed to describe the 
fluctuations of these two observables at a fixed impact parameter value. This 
distribution provides a more precise description of the observable distributions 
in both central and peripheral regions. The effectiveness of the proposed method 
was tested within a realistic BM@N simulation framework for Xe+CsI 
collisions at a beam energy of 3.8$A$ GeV.
\end{abstract}

\maketitle
   
\section{Introduction}
Study on the phase diagram of nuclear matter at high baryonic densities, formed in nucleus-nucleus collisions at beam energies ranging from 2 to 4.5$A$ GeV, is currently being conducted at the BM@N (Baryonic Matter at Nuclotron) experiment~\cite{kapishin2020,senger2022,AFANASIEV2024169532}. BM@N is the first operating fixed-target experiment at the NICA (Nuclotron-based Ion Collider facility) accelerator complex, located at the Joint Institute for Nuclear Research (JINR) in Dubna, Russia. Its experimental program includes measurements of (multi)strange hyperon and hypernuclei yields and studies of the azimuthal asymmetry of the charged particles as a function of the mass of the colliding nuclei, their energy, and centrality.
Various observables sensitive to the transport properties and equation of state of compressed baryonic matter are used to study the phase diagram of nuclear matter; these, in turn, depend on the initial collision geometry of heavy ions \cite{broniowski2002, alice2013, lacey2013,Chen:2024aom}. The impact parameter, characterizing the distance between the centers of the two colliding nuclei, is typically used as a measure to describe the initial geometry. However, this quantity cannot be measured directly in the experiments. Therefore, various centrality determination procedures based on the correlation between an observable quantity and the impact parameter are employed. Typically, the multiplicity of produced charged particles or the spectator energy measured in the forward rapidity region are used as observables \cite{kagamaster2021, loizides2015, parfenov2021, segal2023}.

Using the charged-particle multiplicity as the only centrality observable can introduce an unwanted autocorrelation when the same multiplicity is later employed in fluctuation studies, for instance in the analysis of net-proton number
fluctuations~\cite{luo_volume_2013,na612025}.
In addition, at the beam energies of a few GeV per nucleon the Coulomb deflection of the projectile nucleus by the target one is no longer negligible and must be
incorporated into the description of the collision geometry~\cite{mehndiratta2017}, while enhanced baryon stopping and the
associated energy-conservation constraints require modifications of the
standard Glauber picture~\cite{simak2025}.
The low and discrete multiplicities typical of this energy domain also make it difficult to define narrow centrality classes that keep volume fluctuations under control, a problem that is particularly relevant for event-by-event fluctuation analyses.

Furthermore, in the BM@N experiment, the forward hadron calorimeter (FHCal) exhibits a non-monotonic response as a function of centrality, due to the beam hole in the calorimeter. Thus it is impossible to use this observable alone for unambiguous event classification. These considerations provide a strong motivation for developing an alternative centrality estimator that combines two complementary observables.
The first section of this article provides a brief overview of the main methods for centrality determination in nucleus-nucleus collisions used in current experiments such as STAR, HADES, and ALICE \cite{alice2013, hades2017, star2021}. The second section describes the Bayesian approach for centrality determination using a two-dimensional Gamma distribution. The third section presents the results of applying this method to centrality determination in the BM@N experiment based on simulated data.

\section{Methods for Centrality Determination in Nucleus-Nucleus Collisions}
Currently, there are two main approaches for centrality determination in nucleus-nucleus collisions. The first approach is based on the Glauber model \cite{miller2007}, where the relationship between multiplicity and the impact parameter $b$ is established by matching the measured distribution of produced particle multiplicity with the distribution obtained from the Glauber Monte Carlo method. This method relies entirely on the concepts of participant nucleons $N_{\text{part}}$ and the number of binary nucleon-nucleon collisions $N_{\text{coll}}$, which are used to calculate the number of ``ancestors'' $N_{a} (f)=N_{part} f+(1-f)N_{coll} $, where $f$ characterizes the fraction of inelastic reactions. After calculating the number of ancestors, the number of produced particles can be determined using the convolution with the Negative Binomial Distribution (NBD) $P_{\mu ,k} $: $M_{MC-Gl}(N_a,f,\mu,k) = N_a(f) \times P_{\mu,k}$, where $\mu $ and $k$ are fit parameters chosen to minimize the $\chi^2 / ndf$ value~\cite{loizides2015, alice2013}. This approach has shown good efficiency and has been applied in a number of experiments across a wide energy range, such as STAR, HADES, and ALICE \cite{alice2013, hades2017, kapishin2020}.
The applicability of this approach for centrality determination in the BM@N experiment raises several questions due to the relatively low multiplicity of produced particles, which is explained by the relatively small systems of colliding nuclei and the low collision energies. Consequently, the development of new techniques and models for centrality determination is necessary.
The second approach used for centrality determination is the direct reconstruction method based on Bayes' theorem \cite{das2018, rogly2018}. It allows obtaining information about the impact parameter $b$ using only the measured distribution of an observable quantity $n$ (charged particle multiplicity, transverse energy, etc.), without MC sampling of the entire collision process. Any additive quantity correlated with $b$ can be used as the observable variable $n$. Typically, this is the charged particle multiplicity or transverse energy. This method has been tested on data from ALICE and STAR experiments \cite{das2018, rogly2018}, and has also shown high effectiveness at low energies in the INDRA experiment~\cite{indra2020}. This method has been extended for the case of two observables~\cite{chen2022}. 
In that work, a two-dimensional normal distribution was proposed as the fluctuation kernel for the observables. However, observables such as charged particle multiplicity and transverse energy cannot be described by a normal distribution in the peripheral region; therefore, the Gamma distribution is the best candidate~\cite{rogly2018}. 
A similar situation is observed for spectator energy in central collisions. Thus, 
to describe both central and peripheral collisions, it is necessary to use a 
two-dimensional distribution with a realistic skewness. 
We note, however, that the approach of Ref.~\cite{chen2022} relies on a 
model-independent parameterization of the mean values and covariance matrix elements as exponentials of polynomials in centrality. Already at low polynomial degrees this requires a large number of free parameters (at least 15) and may lead to fit instabilities and possible over-fitting, while practical limitations of the MINUIT fitter package prevent a stable extension to higher degrees, as discussed in that work. Moreover, the smoothly parameterized functions cannot describe the non-monotonic behaviour of the deposited spectator energy in FHCal as a function of centrality (see Fig.~\ref{fig:EnvsImp}), 
which is a distinctive feature of the BM@N experiment. 
These limitations motivate the alternative strategy developed in the present paper. 

\section{The BM@N Experiment}
\begin{figure}[h]
\centering
\includegraphics[width=\hsize]{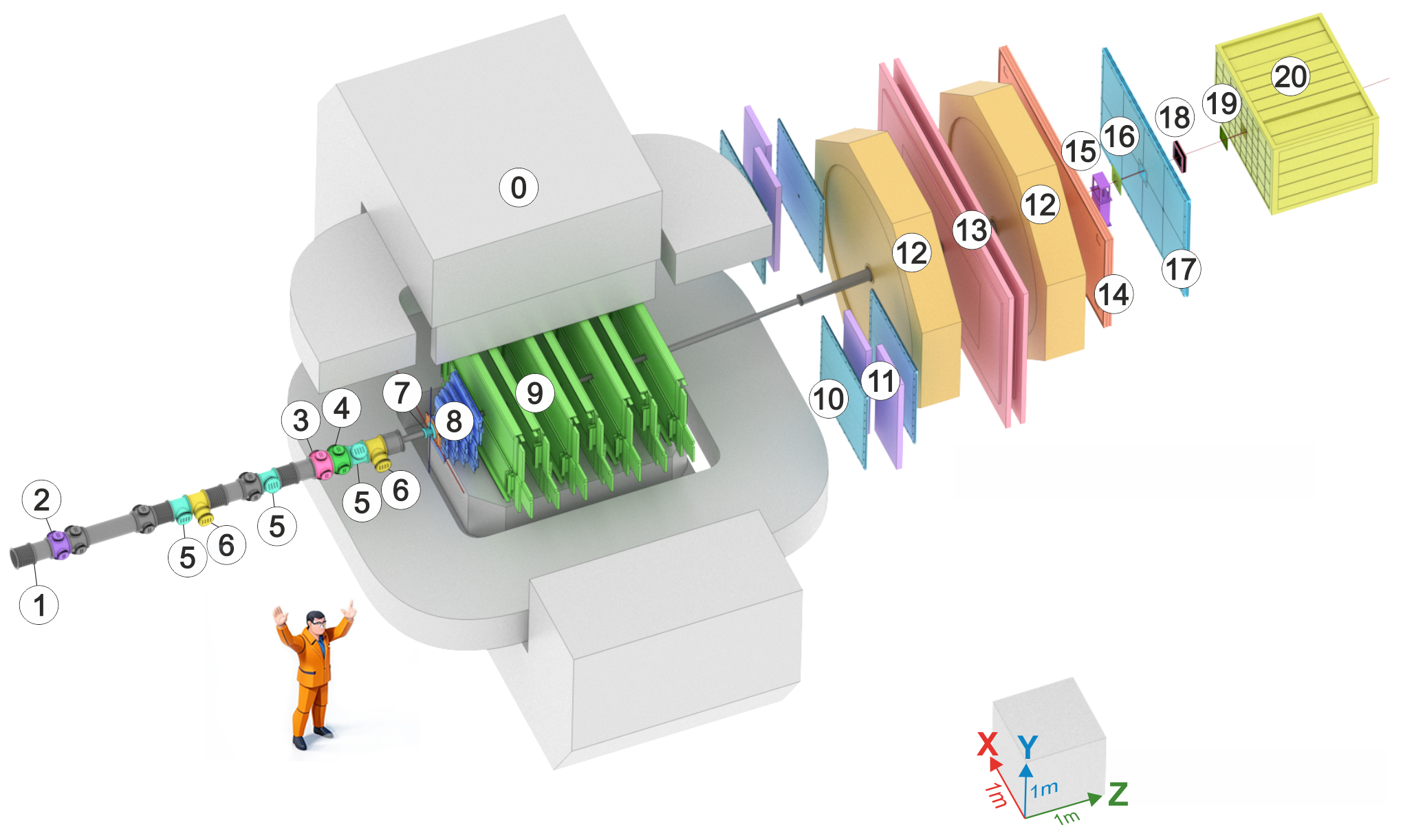}
\caption{Schematic view of the BM@N setup in the 2023 Xe run \cite{AFANASIEV2024169532}.}
\label{fig:setup}
\end{figure}

The results presented in this paper are based on the simulations of the Xe+CsI system at 3.8$A$~GeV, replicating the conditions of the latest BM@N experiment. The event generation was performed with the DCM-QGSM-SMM and UrQMD-AMC models, followed by a detailed simulation of the detector response using GEANT4 within the realistic BM@N geometry. Schematic view of the BM@N setup is shown in Fig.~\ref{fig:setup}.

The trigger system of the BM@N detector is composed of several beam counters and the Barrel Detector (BD) (item 7 in Fig.~\ref{fig:setup}) which aims to provide multiplicity-based event selection. The BD consists of 40 independent plastic scintillator strips arranged in a cylindrical geometry coaxial to the beam, surrounding the target inside the analyzing magnet. Each strip, with dimensions of \(150 \times 7 \times 7\,\text{mm}^3\), has a read-out carried out by a silicon photomultiplier (SiPM) ensuring operation in the strong magnetic field~\cite{afanasiev2024}. To suppress the background from $\delta$-electrons generated by the beam ions, the scintillator array is shielded with an internal cylindrical lead layer. A signal from the BD serves as a part of the Central Collision Trigger (CCT) signal when the hit multiplicity exceeds a programmable threshold, enabling efficient selection of events with varying collision centrality for both online triggering and offline analysis.
The heart of the BM@N setup is its tracking system, designed to precisely follow the paths of the charged particles produced in the collisions. This system is located within a large magnet, whose field bends the trajectories of the particles, allowing their momenta to be measured. The tracking is performed by two main detectors: the Forward Silicon Detector (FSD) (item 8 in Fig. \ref{fig:setup}) and a system of the GEM detectors (item 9 in Fig. \ref{fig:setup}). 

\begin{figure}[htb!]
\centering
\includegraphics[width=0.6\hsize]{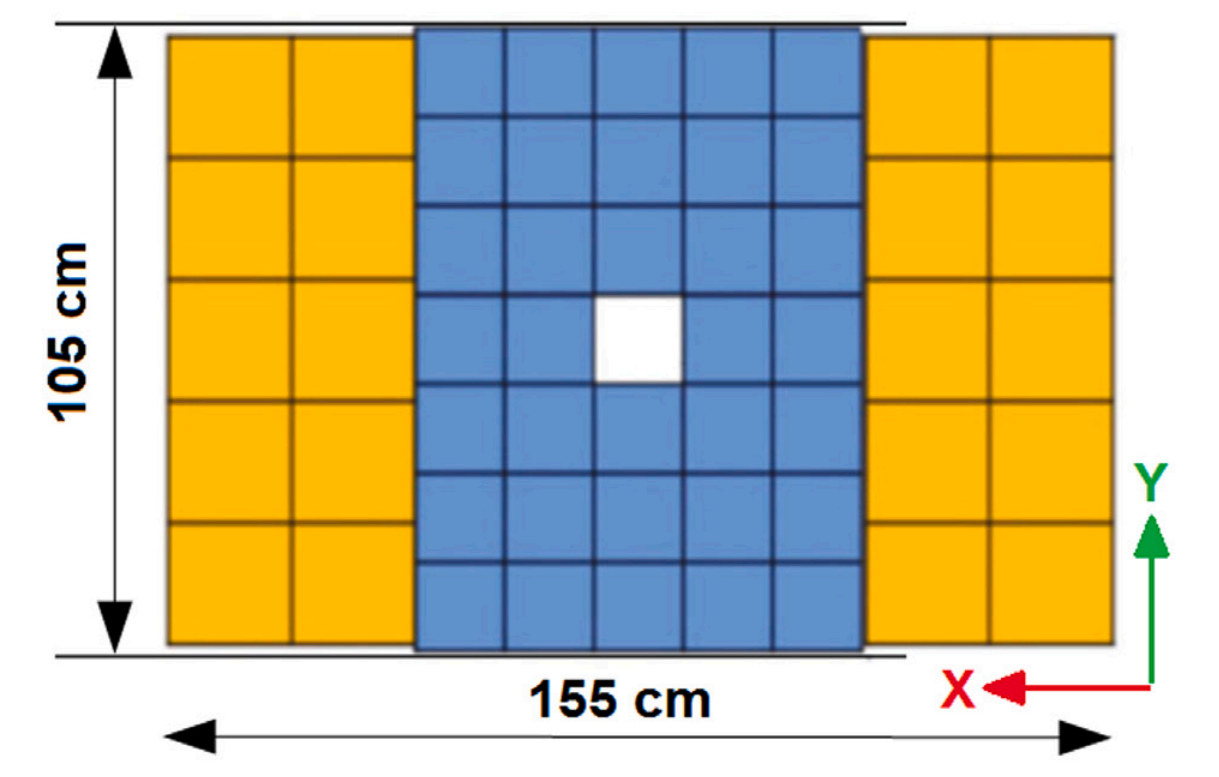}
\caption{Schematic illustration of the FHCal configuration showing the large modules (yellow rectangle) and the missing central module \cite{AFANASIEV2024169532}.}
\label{fig:FHCal_config}
\end{figure}

\begin{figure}[htb!]
\centering
\includegraphics[width=\hsize]{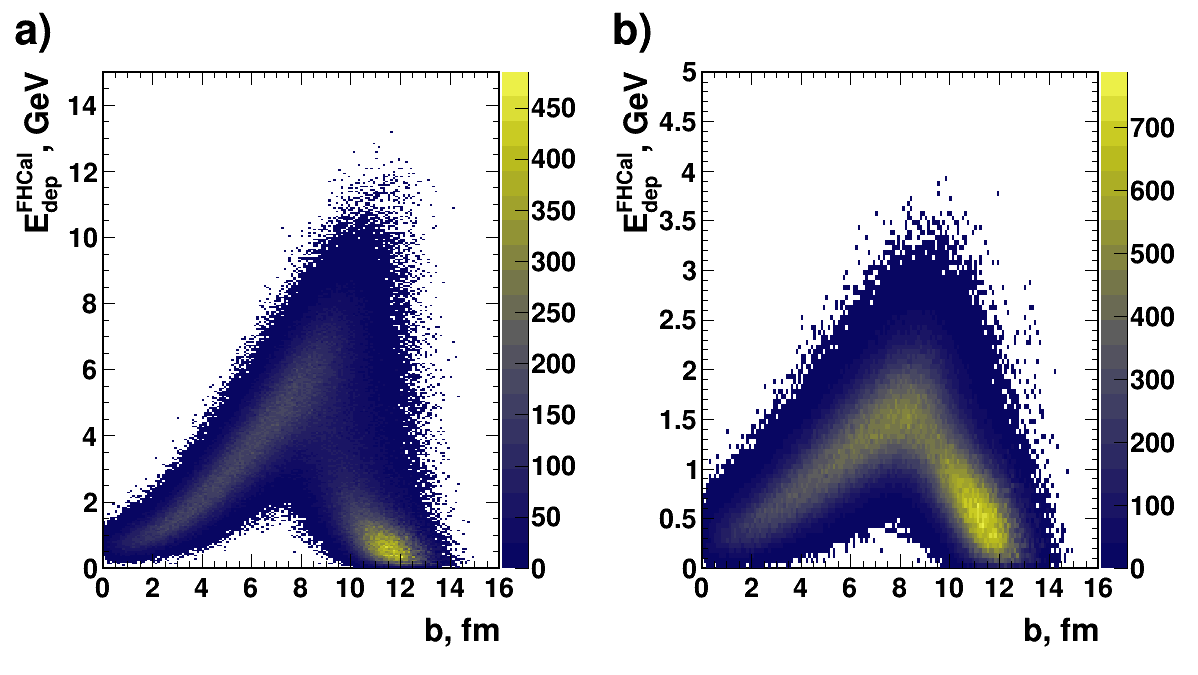}
\caption{Correlation between the deposited spectator energy in FHCal and the impact parameter for Xe+CsI collisions at a beam energy of 3.8 $A$ GeV in DCM-QGSM-SMM model: a) for all modules, b) for large modules \cite{afanasiev2024}.}
\label{fig:EnvsImp}
\end{figure}

The FSD consists of four tracking planes made of high-precision silicon sensors. Each plane is divided into two half-planes (upper and lower) to accommodate the beam vacuum pipe. 
Further downstream, the GEM detectors use gas amplification technology to extend these tracks. By combining the data from both tracking detectors, the experiment can accurately reconstruct the full path and momentum of the charged particles produced in the collision.
At the very end of the BM@N setup, the Forward Hadron Calorimeter (FHCal) (item 20 in Fig. \ref{fig:setup}) is placed \cite{afanasiev2024}. This calorimeter has a modular structure and measures the energy deposition of spectators. The total energy deposition measured by the FHCal can be used to determine the centrality of the collision. 
A distinctive feature of the FHCal in the BM@N is the absence of a central module. Figure~\ref{fig:FHCal_config} illustrates the FHCal configuration, highlighting the large modules and the missing central module.
This leads to a non-monotonic correlation between the energy deposited in the FHCal and the impact parameter.
Figure \ref{fig:EnvsImp} a) presents the correlation between the total energy deposited in the FHCal and the impact parameter. Figure \ref{fig:EnvsImp} b) shows the correlation of the total deposited energy specifically in the large modules with the impact parameter.
The observed difference between these correlations is attributed to the geometry of the BM@N experiment and the positioning of the FHCal in the first physics run Xe+CsI at 3.8$A$ GeV. Due to this configuration, small parts of Xe ion beam and the nuclear fragments from peripheral heavy-ion collisions partially hit the rear part of a single central module. This leads to an anomalous energy deposition in the affected modules that is not associated with the products of the primary collision, thus distorting the expected correlation for peripheral events.
The energy in large modules $E_{dep}^{FHCal}$ is not monotonic relative to the impact parameter and therefore a second observable must be used for the centrality determination. The second observable was chosen to be the number of track hits scaled to the minimum number of hits (4 hits). This observable serves as a proxy for the total charged-particle multiplicity, which is correlated with the number of participant nucleons. The following selection criteria were used for track selection: $0.02 <p_\mathrm{T}<2.0\ \mathrm{GeV}/\textit{c}$, $|DCA_{x,y}|<1\ \mathrm{cm}$ and $0.5<\eta<2.5$, where $p_\mathrm{T}$ and $\eta$ are transverse momentum and pseudorapidity, correspondingly.

\section{Two-Dimensional Direct Reconstruction Method}
\label{sec:TwoDim}
\begin{figure*}[ht!]
\centering
\includegraphics[width=0.95\linewidth]{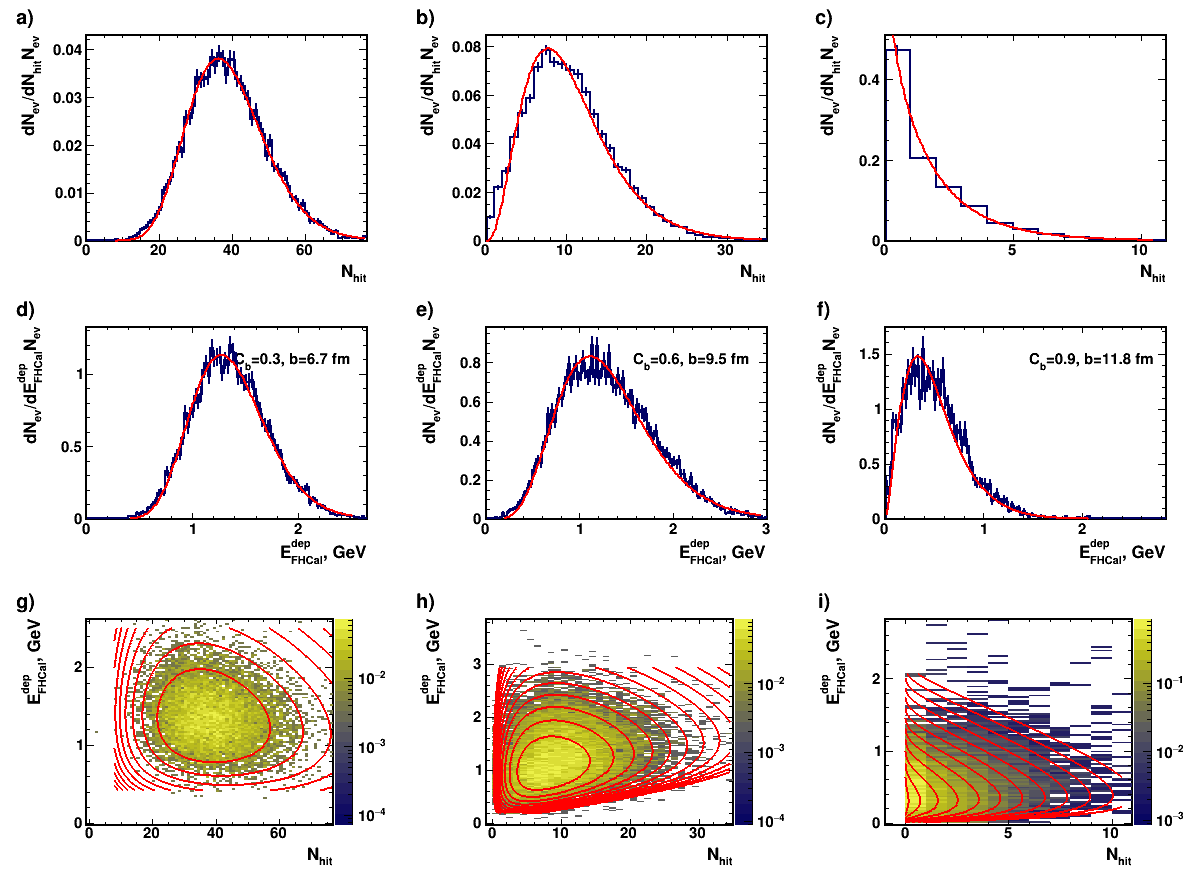}
\caption{Fit results (red curves) of the track hit number distribution (upper panels, a-c) and the deposited energy in the FHCal (middle panels, d-f) for several fixed impact parameter values corresponding to central, semi-central, and peripheral collisions. The lower panels (g-i) show the results of fitting the two-dimensional distribution of observables using function \ref{eq:2DFitFixb}.}
\label{fig:fit_results}
\end{figure*}

\begin{figure}[ht!]
\includegraphics[width=1\linewidth]{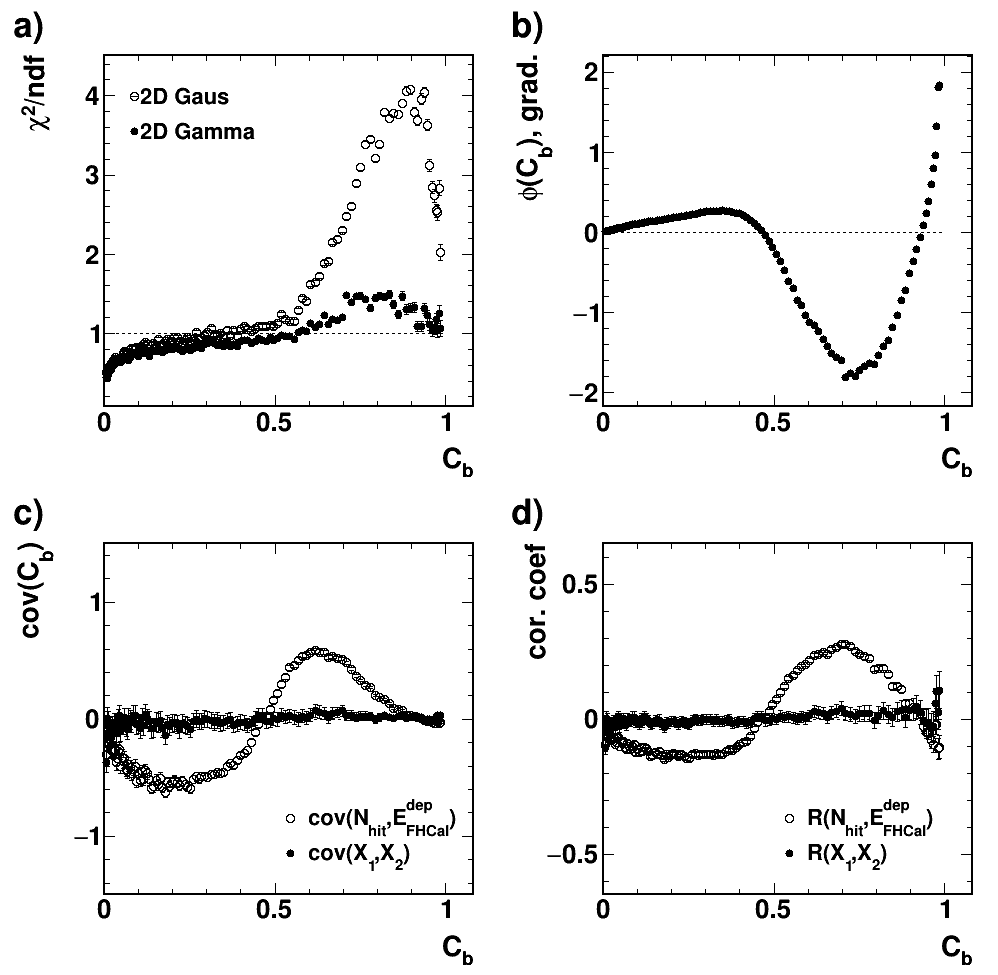}
\caption{(a) $\chi^2$/ndf vs centrality for fits with a bivariate normal distribution and the proposed approach. (b) Rotation angle $\phi$ vs centrality. (c, d) Covariance and correlation coefficient for $E_{\text{sp}}$ and $N_{\text{hit}}$ (and for the new variables $X_{1}$ and $X_{2}$) as functions of centrality. The residual correlation in the rotated coordinates is negligible ($|\rho| < 0.01$) for all centralities.}
\label{fig:Chi2NDF_CoV}
\end{figure}
One of the main ideas of the proposed new method is that the function $P(E_{\text{sp}}, N_{\text{hits}}|b)$, characterizing the probability density distribution of spectator energy $E_{\text{sp}}$, deposited in the forward hadron calorimeter, and the number of charged particle hits $N_{\text{hits}}$ to have a specific $b$, can be approximated by a two-dimensional Gamma distribution $\Gamma_2(E_{\text{sp}}, N_{\text{hits}}|b)$. However, there is currently no explicit analytical expression for this distribution \cite{tashkandy2018}, and all proposed variants are applicable only within a narrow range of parameter values. In the case where the two random variables $X_1$ and $X_2$ are independent, the two-dimensional distribution for these quantities will be equal to the product of their distributions at fixed $b$:
\begin{equation}\label{eq:prob_gamma2D}
P(X_1, X_2|b) = \prod_{i=1}^2 \frac{X_i^{k_i(b)-1} e^{-X_i/\theta_i(b)}}{\Gamma(k_i(b)) \theta_i(b)^{k_i(b)}}
\end{equation}
where $k_i(b)$ and $\theta_i(b)$ are positive parameters that generally depend on $b$ and are defined as follows:
\begin{equation}\label{eq:gamma2d_components}
k_i(b) = \frac{\langle X_i(b) \rangle^2}{\sigma_{X_i}^2(b)}, \quad \theta_i(b) = \frac{\sigma_{X_i}^2(b)}{\langle X_i(b) \rangle}
\end{equation}
Here $\sigma_{X_i}^2(b)$ and $\langle X_i(b) \rangle$ denote the variance and mean value of the random variable, respectively.

At a fixed impact parameter, the correlation between $E_{sp}$ and $N_{hit}$ is dominated by a linear contribution, since at low energies the multiplicity of produced particles is proportional to the number of participant nucleons ($N_{\text{hits}} \propto N_{\text{part}}$), whereas the energy deposited in the FHCal is proportional to the number of spectator nucleons ($E_{\text{sp}} \propto N_{\text{sp}} = 2A - N_{\text{part}}$) and participant nucleons. Our key assumption is that after removing this dominant linear correlation, the remaining fluctuations in the rotated coordinates become sufficiently weakly correlated to be approximated by independent Gamma distributions. The new variables $X_1$ and $X_2$ can be obtained from $E_{sp}$ and $N_{hit}$ by rotating the coordinate axes through an angle:
\begin{equation}
\phi(b)=\frac{1}{2}\arctan\!\left(\frac{2\,\mathrm{Cov}(E_{\mathrm{sp}},N_{\mathrm{hit}})}{\sigma_{E_{\mathrm{sp}}}^2(b)-\sigma_{N_{\mathrm{hit}}}^2(b)}\right)
\label{eq:rotANG}
\end{equation}

Where $\text{Cov}(E_{sp} ,N_{hit})$ is the covariance of the number of track hits of particles and the spectator energy at the fixed impact parameter. The rotation defined by Eq.~(\ref{eq:rotANG}) can be obtained by analogy
with the principal‑axis transformation used in Principal Component Analysis
(PCA)~\cite{jolliffe2002} and with standard algorithms for generating
correlated random variables with a prescribed covariance
matrix~\cite{gentle2009}. If the units of the observables are changed,
the angle $\phi$ is recalculated accordingly, and the procedure remains
internally consistent. Since the covariance and variances depend on $b$, the rotation angle $\phi$ is also a function of centrality $\phi = \phi(c_b)$. It can easily be shown that for the new variables $X_1$, $X_2$, the condition $\text{Cov}(X_1, X_2) = 0$ holds. We stress that for non‑Gaussian variables a vanishing covariance does not
guarantee statistical independence; hence Eq.~(\ref{eq:2DFitFixb}) is an
approximation, whose adequacy is verified a posteriori through
the $\chi^2/\mathrm{ndf}$ and residual‑correlation studies presented in
Fig.~\ref{fig:Chi2NDF_CoV}.
The relationship between the quantities $X_1$, $X_2$ and $N_{\text{hits}}$, $E_{\text{sp}}$ is determined through the rotation matrix, and the mean values and variances are defined as:
\begin{equation}\label{eq:rot_matrix}
\begin{pmatrix}
\langle X_1(b) \rangle\\
\langle X_2 (b)\rangle
\end{pmatrix}
=
\begin{pmatrix}
\cos(\phi(b)) & -\sin(\phi(b)) \\
\sin(\phi(b)) & \cos(\phi(b))
\end{pmatrix}
\begin{pmatrix}
\langle N_{\text{hits}}(b) \rangle\\
\langle E_{\text{sp}}(b) \rangle
\end{pmatrix}
\end{equation}
\begin{equation}\label{eq:rot_var_x1}
\begin{gathered} 
\sigma_{X_1}^2(b) = \cos^2(\phi(b)) \sigma_{N_{\text{hits}}}^2(b) + 
\sin^2(\phi(b)) \sigma_{E_{\text{sp}}}^2(b) -\\ \sin(2\phi(b)) \text{Cov}(N_{\text{hits}}, E_{\text{sp}})
\end{gathered} 
\end{equation}
\begin{equation}\label{eq:rot_var_x2}
\begin{gathered} 
\sigma_{X_2}^2(b) = \sin^2(\phi(b)) \sigma_{N_{\text{hits}}}^2(b) + \cos^2(\phi(b)) \sigma_{E_{\text{sp}}}^2(b) +\\ \sin(2\phi(b)) \text{Cov}(N_{\text{hits}}, E_{\text{sp}})
\end{gathered} 
\end{equation}

Thus, the distribution for the observables $N_{\text{hits}}$, $E_{\text{sp}}$ in the rotated coordinates can be approximated by the two-dimensional Gamma distribution as follows:
\begin{equation}
\begin{gathered} 
P(N_{\text{hits}}, E_{\text{sp}}|b) \approx P(X_{1}(N_{\text{hits}},E_{\text{sp}}), X_{2}(N_{\text{hits}},E_{\text{sp}})|b) =\\ \Gamma(k_1(b), \theta_1(b)) \cdot \Gamma(k_2(b), \theta_2(b))
\label{eq:2DFitFixb}
\end{gathered} 
\end{equation}
where $\Gamma(k, \theta)$ is the Gamma distribution. The distribution parameters, $k_i(b)$ and $\theta_i(b)$, can be calculated using expressions (\ref{eq:gamma2d_components}-\ref{eq:rot_var_x2}). Figure \ref{fig:fit_results} shows the results of fitting the distribution of the number of track hits and the energy in the FHCal calorimeter in DCM-QGSM-SMM model for Xe+CsI collisions at a beam energy of 3.8$A$ GeV. The obtained results show that both observables can be well described by a Gamma distribution at a fixed value of the impact parameter. The lower panels show the results of fitting the two-dimensional distribution of observables using function \eqref{eq:2DFitFixb}. The centrality dependence of the $\chi^2$/ndf parameter obtained from fitting the two-dimensional distribution at fixed impact parameter is shown in Fig. \ref{fig:Chi2NDF_CoV}(a) for both the proposed approach and a two-dimensional normal distribution. The comparison demonstrates better description achieved with the proposed two-dimensional Gamma distribution approach. Fig. \ref{fig:Chi2NDF_CoV}(b) presents the rotation angle $\phi$, calculated via Eq.~\ref{eq:rotANG}, as a function of centrality. Fig. \ref{fig:Chi2NDF_CoV}(c, d) shows the dependence of the covariance and the Pearson correlation coefficient between $E_{sp}$ and $N_{hit}$ on centrality, as well as for the new variables $X_1$ and $X_2$. The results displayed in Fig. \ref{fig:Chi2NDF_CoV}(c,d) confirm that the axis rotation by $\phi$ successfully removes the dominant linear correlation between the observables.
\begin{figure}[ht!]
\includegraphics[width=1\linewidth]{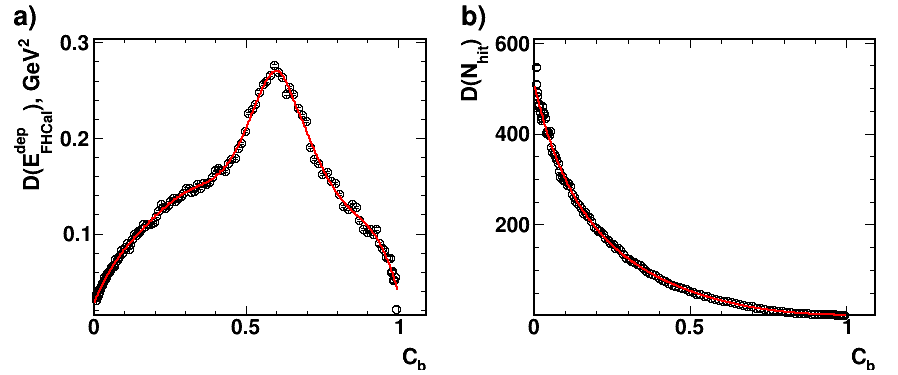}
\caption{The left plot shows dependence of the variance of the deposited spectator energy in FHCal and the right plot shows variance of the number of track hits, as well as their covariance on centrality. Open symbols represent the data from the DCM-QGSM-SMM model, and the red line shows the fit with a polynomial function.}
\label{fig:mean_var}
\end{figure}

\begin{figure}[htb!]
\includegraphics[width=1\linewidth]{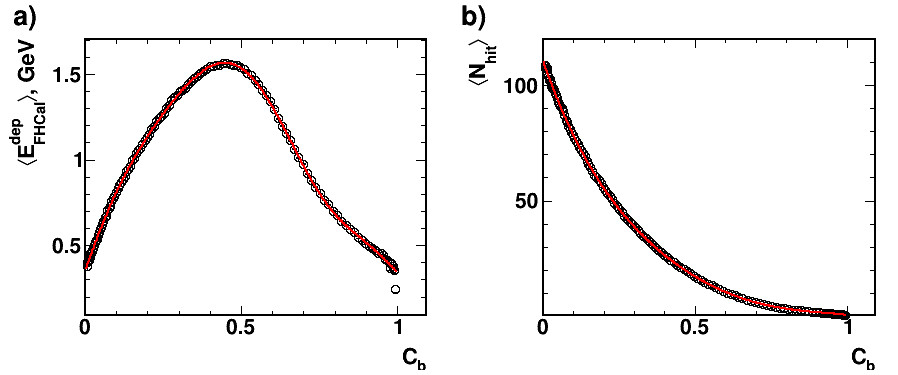}
\caption{The left plot shows distribution of the mean values of the deposited spectator energy and the right plot shows number of track hits as a function of centrality. The results from the DCM-QGSM-SMM model (open symbols) are fitted with a polynomial function, shown as the red curve.}
\label{fig:mean}
\end{figure}

The distributions of the mean value and variance of the observables from the model are approximated by polynomials $f(c_b)$, where $c_b$ is the true centrality or $b$-centrality, defined as follows 

\begin{equation}
c_b =  \int_0^b P(b') db'= \frac{1}{{\sigma_{\text{inel}}}}\int_0^b2\pi bP_{inel}(b') db'.
\label{eq:CB}
\end{equation}

Here $P(b)$ is the probability distribution of the impact parameter, $\sigma_{\text{inel}}$ is the cross-section for inelastic nucleus-nucleus interaction, and $P_{\text{inel}}(b)$ is the probability of an inelastic collision. 
The results of fitting the dependence of the variance of the observables in the model, as well as their covariance, on centrality are presented in Fig. \ref{fig:mean_var}. The dependencies of the mean values of the observables on centrality are depicted in Fig. \ref{fig:mean}. 

The observed probability density distribution of spectator energy and the number of charged particle hits $P(E_{\text{sp}}, N_{\text{hits}})$ is related to the probability density at a fixed value of the impact parameter $P(E_{\text{sp}}, N_{\text{hits}}|b)$ by the following expression:
\begin{equation}
\begin{gathered} 
P(E_{\text{sp}}, N_{\text{hits}}) = \int_0^\infty P(E_{\text{sp}}, N_{\text{hits}}|b) P(b) db=
\\\int_0^1 P(E_{\text{sp}}, N_{\text{hits}}|c_{b}) dc_{b}
\label{eq:2DFit}
\end{gathered} 
\end{equation}
where $P(E_{\text{sp}}, N_{\text{hits}}|c_{b})$ is derived from $P(E_{\text{sp}}, N_{\text{hits}}|b)$ through the use of the substitution described by the equation~\eqref{eq:CB}. The full integral $P(E_{\text{sp}}, N_{\text{hits}})$ can be used to fit the two-dimensional distribution of energy and number of hits in the experiment, assuming that the mean values and variances of the observables are proportional to the values obtained from fully reconstructed model data:
\begin{equation}
\langle E_{\text{sp}}(b) \rangle = \alpha_E \langle E_{\text{sp}}^{\text{MC}}(b)\rangle,
\langle N_{\text{hits}}(b) \rangle = \alpha_N \langle N_{\text{hits}}^{\text{MC}}(b)\rangle
\end{equation}
\begin{equation}
\sigma_{E_{\text{sp}}}^2(b) = \alpha_E^{2}\sigma_{E_{\text{sp}}^{\text{MC}}}^2(b)+\alpha_E\beta_E\langle E_{\text{sp}}^{\text{MC}}(b) \rangle \end{equation}
\begin{equation}
\sigma_{N_{\text{hits}}}^2(b) = \alpha_N^{2}\sigma_{N_{\text{hits}}^{\text{MC}}}^2(b)+\alpha_N\beta_N\langle N_{\text{hits}}^{\text{MC}}(b) \rangle
\end{equation}
where $\alpha_E, \alpha_N, \beta_E, \beta_N$ are fit parameters. The coefficients introduced in this way allow accounting for the difference between experimental and model data in describing the mean values and variances of the observables. Such differences are due to variations in the gain coefficients of individual calorimeter readout channels, calibration errors, and temperature effects.

Another important factor that must also be taken into account is the efficiency of the trigger system, as it will cause a fraction of events to be missed, which will affect the normalization of the distributions. To account for this effect, another parameter $\epsilon=\text{N}_{Raw}^{Ev}/\text{N}_{Total}^{Ev}$ 
is introduced. Here, $\mathrm{N}_{Raw}^{Ev}$ denotes the number of events recorded experimentally, 
while $\mathrm{N}_{Total}^{Ev}$ corresponds to the number of events that would be registered with 
a perfect (100\% efficient) trigger. Consequently, the parameter $\epsilon$ represents the total 
event‑detection efficiency. 
In the present proof-of-principle study, $\epsilon$ is treated as a free parameter in the fit; 
for application to the real experimental data, it should be constrained by an independent efficiency 
measurement. In the current implementation, we assume it to be a global constant independent of centrality, characterizing the overall fraction of inelastic events that pass the trigger and selection criteria. Thus, the considered method includes the following parameters - $\alpha_E, \alpha_N, \beta_E, \beta_N, \epsilon$, which can be determined by fitting the two-dimensional distribution. Once the fit parameters are determined, the probability distribution of the impact parameter for a fixed range of the number of hits($N_{1},\ N_{2}$) and the energy of spectators($E_{1},\ E_{2}$) can be determined from Bayes' theorem:
\begin{equation}
\begin{gathered} 
P(b|E_{1}<E_{\text{sp}}<E_{2},N_{1}<N_{\text{hits}}<N_{2})=
\\P_{inel}\frac{\int_{E_{1}}^{E_{2} }\int_{N_{1}}^{N_{2}}P(E,N|c_{b})dE_{\text{sp}}dN_{\text{hits}} }{\int_{E_{1}}^{E_{2}} \int_{N_{1}}^{N_{2}}\int_{0}^{1}P(E,N|c_{b})dE_{\text{sp}}dN_{\text{hits}}dc_{b}}
\end{gathered} 
\label{eq:bayes_2d}
\end{equation}
The proposed approach can also be applied for centrality determination using a single observable, such as $N_{\text{hits}}$. An advantage of this method compared to the previous approaches \cite{das2018, rogly2018} is the smaller number of fit parameters $\alpha_N, \beta_{N}, \epsilon$, and the possibility to account for the trigger efficiency. The fit function can be written as follows:
\begin{equation}
\begin{gathered} 
F(N_{\text{hits}}) =\int_0^1 P(N_{\text{hits}}|c_{b}) dc_{b}/\epsilon
\label{eq:1DFit}
\end{gathered} 
\end{equation}
where $P(N_{\text{hits}}|c_{b})$ is a Gamma function with parameters $\theta(c_{b})=\sigma_{N_{\text{hits}}}^2(c_{b})/\langle N_{\text{hits}}(c_{b}) \rangle$ and $k(c_{b})=\langle N_{\text{hits}}(c_{b}) \rangle^2/\sigma_{N_{\text{hits}}}^2(c_{b})$. With the fitted parameters $\alpha_N$, $\beta_N$, and $\epsilon$ obtained, the posterior probability density $P(b)$ of the impact parameter for a centrality class defined by track-hit number cuts $(N_1, N_2)$ is calculated via Bayes' theorem.
\begin{eqnarray}
	\label{eq:Bayes}
	P(b|N_1 < N_{\text{hits}} < N_2) = P(b)\frac{\int \limits_{N_1}^{N_2} P(N_{\text{hits}}|b) dN_{\text{hits}}}{\int \limits_{N_1}^{N_2} P(N_{\text{hits}}) dN_{\text{hits}}}.
\end{eqnarray}

\section{Centrality Determination in the BM@N Experiment}
The results presented in this section are based on simulations of the Xe+CsI system at $3.8A$~GeV, replicating the conditions of the latest BM@N experiment. We analyze one- and two-dimensional distributions of charged particle multiplicity and spectator energy. The event generation was performed with two models, \textsc{DCM-QGSM-SMM}~\cite{baznat2020} and \textsc{UrQMD-AMC}~\cite{Svetlichny:2025ocm}, followed by a detailed simulation of the detector response using \textsc{GEANT4}~\cite{Geant4} within the realistic BM@N geometry.

The DCM-QGSM-SMM model is based on the Dubna Cascade Model (DCM-QGSM) and the Statistical Fragmentation Model (SMM). The model is intended for and used to generate particle-nucleus and nucleus-nucleus collisions in a wide energy range. The DCM-QGSM-SMM model can also simulate the production of both light particles and nuclear fragments, and hyperfragments. 

The second model applied in this work is the hybrid UrQMD-AMC approach. It combines the Ultrarelativistic Quantum Molecular Dynamics (UrQMD) transport code with the Ablation Monte Carlo (AMC) framework for clustering and statistical decay of spectator matter. This model was specifically utilized to test and validate the method for determining collision centrality based on the production of forward spectator neutrons, protons, and light nuclear fragments.

After obtaining the dependencies of the means, variances, and covariance, the probability density distribution for the specified observables can be obtained; the resulting function \eqref{eq:2DFit} was used to fit the simulation data. The centrality determination method was further validated on simulated data from the DCM-QGSM-SMM and UrQMD-AMC event generators, incorporating a realistic trigger simulation. To emulate the experimental trigger conditions, events were selected with a barrel detector multiplicity $N_{\text{barrel}} \geq 4$ and more than two reconstructed vertex tracks. The fit parameters $\alpha_E$, $\alpha_N$, $\beta_E$, and $\beta_N$ obtained for the pure model and for the trigger-simulated data are listed in Tables \ref{tab:fit_parametersDCM} and \ref{tab:fit_parametersUrQMD}. 

\begin{table}[h]
\centering
\caption{Fit parameters for the DCM-QGSM-SMM model.}
\label{tab:fit_parametersDCM}
\begin{tabular}{lccc}
\toprule
\textbf{Fit par.} & \textbf{Model} & \textbf{With sim. trig.} \\
\midrule
$\alpha_E$ & $1.001 \pm 0.002$ & $0.988 \pm 0.003$ \\
$\beta_E$ & $-0.009 \pm 0.001$ & $-0.004 \pm 0.002$ \\
$\alpha_N$ & $0.972 \pm 0.002$ & $1.003 \pm 0.004$ \\
$\beta_N$ & $0.013 \pm 0.007$ & $-0.035 \pm 0.012$ \\
$\epsilon$ & $1.004 \pm 0.004$ & $0.65\pm 0.02$ \\
\bottomrule
\end{tabular}
\end{table}

\begin{table}[h]
\centering
\caption{Fit parameters for the UrQMD-AMC model.}
\label{tab:fit_parametersUrQMD}
\begin{tabular}{lccc}
\toprule
\textbf{Fit par.} & \textbf{Model} & \textbf{With sim. trig.} \\
\midrule
$\alpha_E$ & $1.002 \pm 0.002$ & $0.991 \pm 0.003$ \\
$\beta_E$ & $-0.006 \pm 0.001$ & $-0.003 \pm 0.001$ \\
$\alpha_N$ & $0.979 \pm 0.002$ & $1.004 \pm 0.003$ \\
$\beta_N$ & $0.011 \pm 0.007$ & $-0.037 \pm 0.008$ \\
$\epsilon$ & $1.002 \pm 0.005$ & $0.64 \pm 0.02$ \\
\bottomrule
\end{tabular}
\end{table}

Since the estimator is constructed from the mean and variance of observables from the same simulated sample, the resulting parameters $\alpha_E$ and $\alpha_N$ are close to 1, while $\beta_E$ and $\beta_N$ are close to zero, as expected for a self-consistent calibration. This confirms that the centrality determination procedure functions correctly within both theoretical frameworks under realistic experimental conditions.

\begin{figure}[h]
\includegraphics[width=1\linewidth]{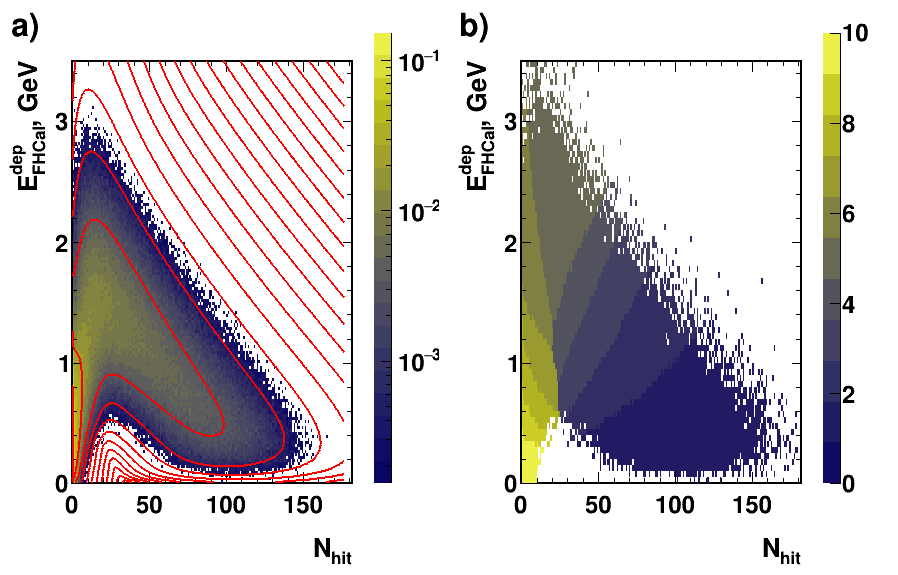}
\caption{(Left) Results of fitting the two-dimensional distribution of the number of hits and deposited energy of the spectators in the FHCal calorimeter in DCM-QGSM-SMM model. The fit results are represented by the red line. (Right) Centrality classes for the two-dimensional distribution. The classes are defined using constrained k-means, where each class contains exactly 10\% of the total number of observed events.}
\label{fig:2d_fit}
\end{figure}

\begin{figure}[h]
\includegraphics[width=1\linewidth]{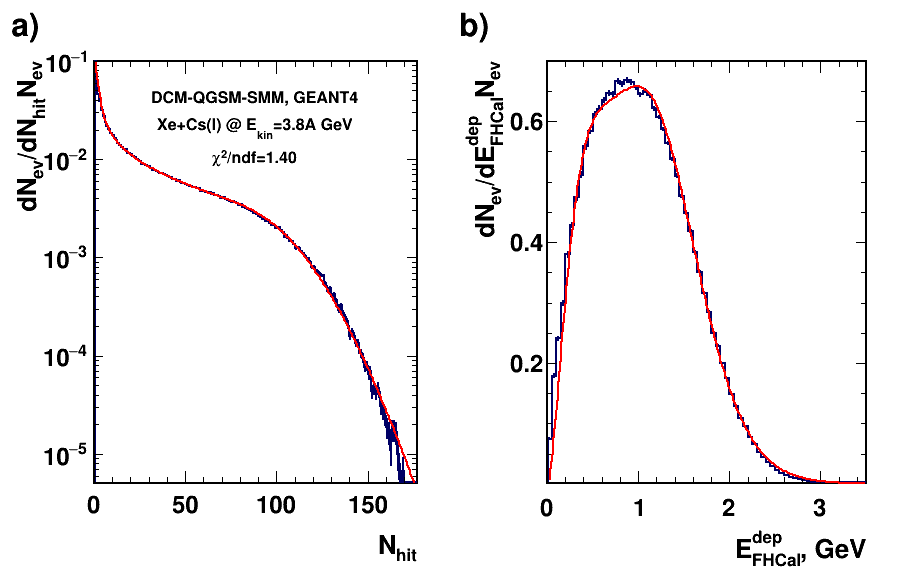}
\caption{The left plot shows the two-dimensional fit results projected onto the hit number axis, the right plot shows the projection onto the FHCal energy axis. The fit results are represented by the red line, and the DCM-QGSM-SMM model results are shown as the blue line.}
\label{fig:2d_fit_proj}
\end{figure}

Figure \ref{fig:2d_fit} shows the results of fitting such a function to the two-dimensional distribution of measured spectator energy and the number of charged particle hits in simulation data for collisions of the Xe+CsI reaction at a xenon nucleus energy of 3.8 $A$ GeV in the BM@N experiment. The results of the two-dimensional fit projected onto the axes are presented in Figure \ref{fig:2d_fit_proj}. 

The obtained results demonstrate good agreement between the fit and the model data. 

After fitting, the obtained two-dimensional distribution was divided into 10 centrality classes using the k-means constrained method \cite{ConstrainedKMeans} (Figure \ref{fig:2d_fit}~b). Unlike standard k-means, which minimizes the within-cluster distance without constraints, the constrained variant additionally controls the fraction of events in each cluster, which allows to account for the required percentage of events within a specific centrality class. Particularly, we impose that each class contains exactly 10\% of the total number of observed events. This is equivalent to defining classes as 10 percentiles of the event distribution in the plane of observables $(E_{\text{sp}}, N_{\text{hits}})$. The k-means-constrained library was used for implementation [https://pypi.org/project/k-means-constrained/]. The cluster centroids were set as the mean values of the observables for the corresponding classes. For this purpose, the previously obtained dependencies of the means on centrality were used. The described method leads to a more reliable and stable event classification across centrality intervals.
\begin{figure}[ht!]
\includegraphics[width=1\linewidth]{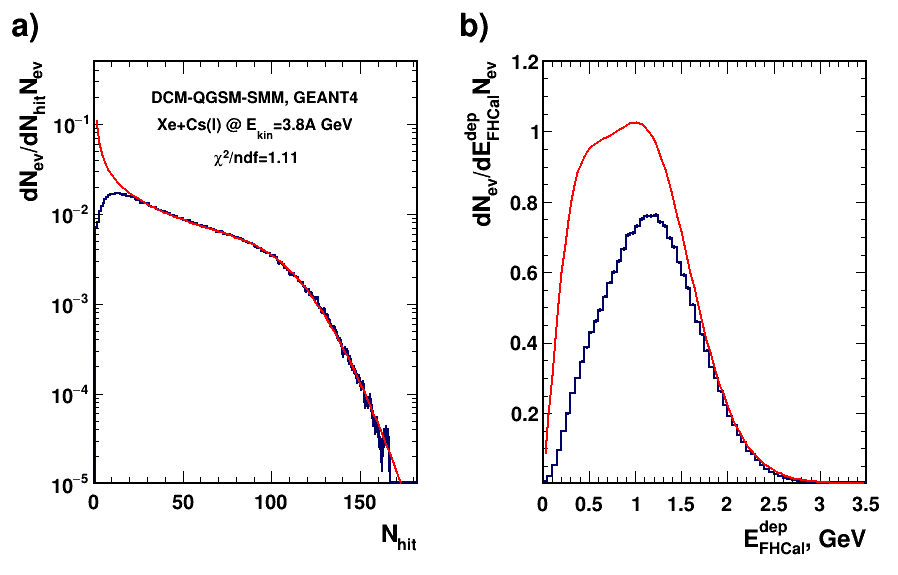}
\caption{The two-dimensional fit results for simulated data including trigger effects are shown in projection: (left) onto the track-hit multiplicity ($N_{\text{hit}}$) axis and (right) onto the spectator energy ($E_{\text{sp}}$) axis. The fit results are represented by the red line. While the fit quality is slightly reduced compared to the pure model case, particularly in the tail of the $E_{\text{sp}}$ distribution, the method remains stable and captures the overall shape.}
\label{fig:2d_fit_projTrig}
\end{figure}

\begin{figure}[ht!]
\includegraphics[width=1\linewidth]{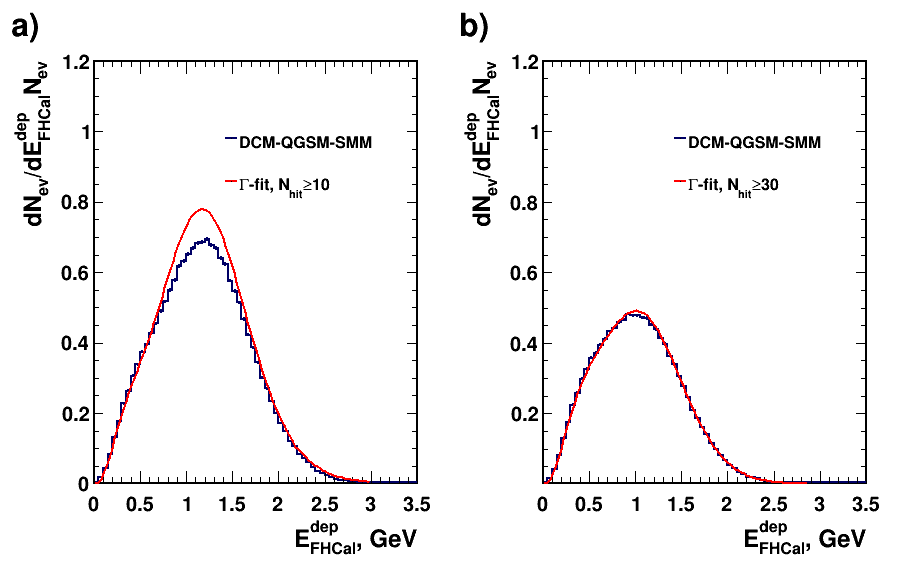}
\caption{ The deposited energy of the spectators distributions below (left) and above (right) the anchor point $N_{\text{hits}}=25$.}
\label{fig:EnFit}
\end{figure}

Figure \ref{fig:2d_fit_projTrig} presents the fit results obtained from simulated data that included a realistic trigger selection. The slight discrepancy visible in the $N_{\text{hits}}$ projection is confined to the low‑multiplicity region ($N_{\text{hits}} \lesssim 25$, i.e.\ peripheral collisions), while for $E_{\text{sp}}$ the fit‑model difference extends over a broader interval. This behaviour reflects the non‑monotonic dependence of the deposited spectator energy on centrality, which spreads the deviation across a wider range of the distribution compared to the monotonically falling $N_{\text{hits}}$ spectrum. To demonstrate this explicitly, we have compared the model and the fit results for a restricted subsample with $N_{\text{hits}} \ge 25$, i.e.\ below the ``anchor point'' of the non‑monotonic behaviour (see Fig.~\ref{fig:EnFit}).

The method demonstrates stable fitting performance and is therefore suitable for application in actual experimental conditions. The overall event detection efficiency extracted directly from the simulation is 
$\epsilon_{\text{model}} = 0.64$ for both the DCM‑QGSM‑SMM and UrQMD‑AMC models. 
The values obtained from the fit, $\epsilon_{\text{fit}} = 0.65 \pm 0.02$ for 
DCM‑QGSM‑SMM (Table~\ref{tab:fit_parametersDCM}) and $\epsilon_{\text{fit}} = 0.64 \pm 0.02$ 
for UrQMD‑AMC (Table~\ref{tab:fit_parametersUrQMD}), are consistent with the simulation 
input. This agreement demonstrates that the Bayesian framework can correctly recover the 
known efficiency from the simulated data. For real experimental data, $\epsilon$ should be 
determined or at least constrained by an independent efficiency study; under such conditions 
the proposed method may itself serve as a complementary cross‑check of the total event‑detection 
efficiency.

\begin{figure}[ht]
\centering
\includegraphics[width=1\linewidth]{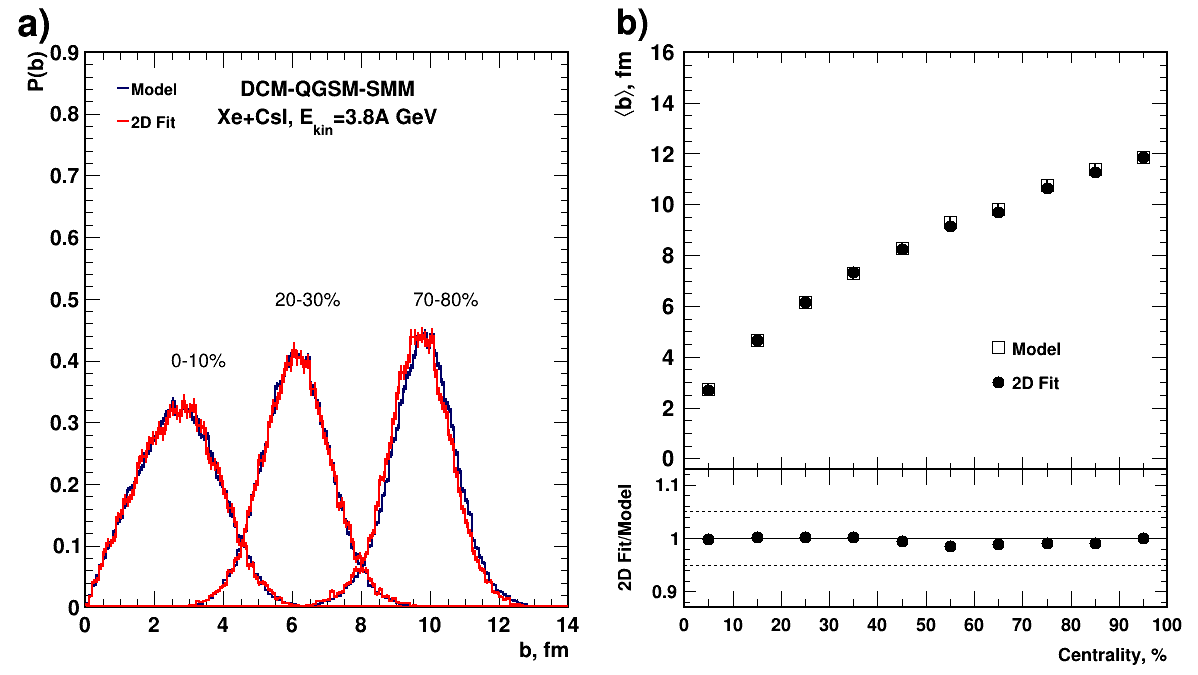}
\caption{(Left) Distribution of the impact parameter obtained from the two-dimensional fit (red line) and from the model (blue line) for three centrality classes: 0–10\% (most central), 20–30\% (mid-central), and 70–80\% (peripheral). (Right) Mean impact parameter $\langle b \rangle$ as a function of centrality percentile, derived from the model (open circles) and the fit procedure (filled squares). The error bars represent the statistical errors of the mean of the distributions. The bottom panel shows the ratio of the fit results to the model values.}
\label{fig:ImpDist}
\end{figure}

Figure \ref{fig:ImpDist} (a) shows the distributions of the impact parameter for three classes of centrality. The impact parameter distributions from the model and from the fit were obtained using the same centrality classes, which are shown in Figure \ref{fig:2d_fit} (b). The red lines indicate the distributions obtained using the Bayesian approach, where the blue lines indicate the results from the DCM-QGSM-SMM model. The results of comparing the dependence of the mean value of the impact parameter on centrality are shown in Figure \ref{fig:ImpDist} (b). The black closed circles represent the results from the fit, and the open black squares represent the data from the model. The obtained distributions agree within 2\%.  

\begin{figure}[!ht]
\centering
\includegraphics[width=1\linewidth]{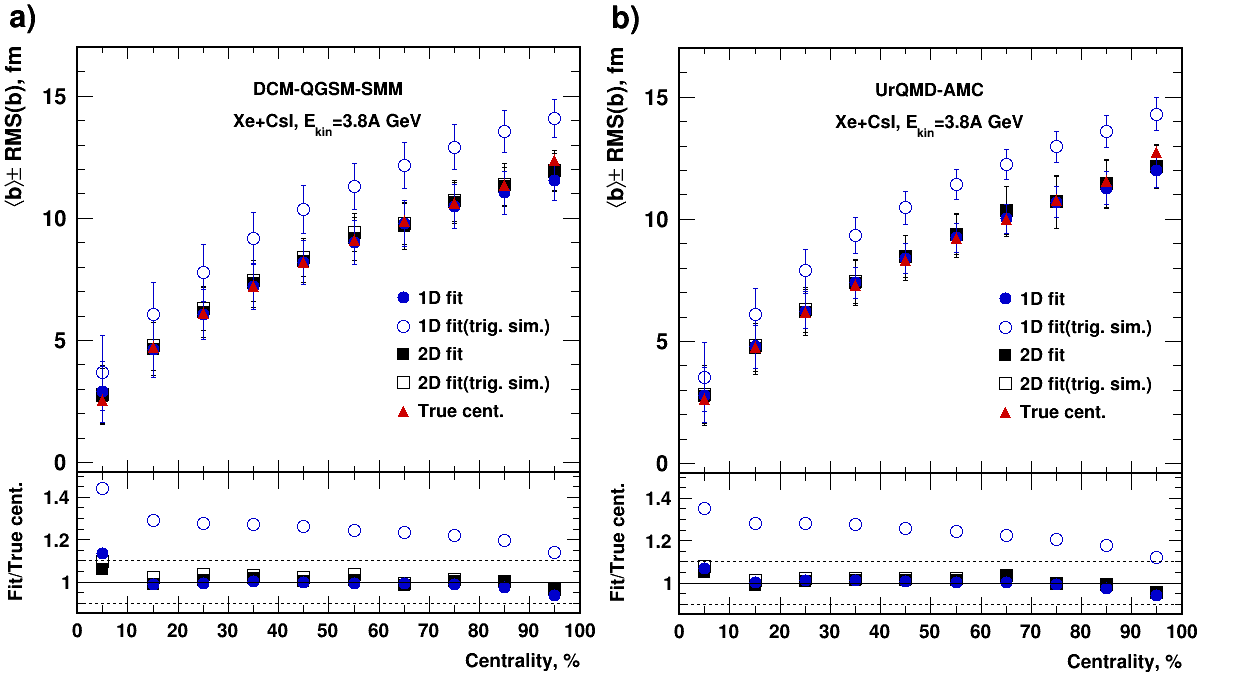}
\caption{Centrality dependence of the mean impact parameter reconstructed using different methods and model datasets. Filled markers correspond to fits of pure model data (left: DCM‑QGSM‑SMM; right: UrQMD‑AMC), while open markers represent fits of data with simulated trigger selection. Results of the one‑dimensional fit to the $N_{\text{hit}}$ distribution using the method from Ref.~\cite{rogly2018} are shown with circles. Triangles show the mean impact parameter for centrality classes defined from the model's impact parameter distribution (true centrality). The lower panel shows the ratio of the mean values obtained with the different methods to the mean $b$ from the true centrality classes.}
\label{fig:CompMeanImp}
\end{figure}
The results show that the proposed approach accurately describes the fluctuations of the observables and allows for accurate reconstruction of the impact parameter distribution.

\begin{figure}[ht!]
\centering
\includegraphics[width=1\linewidth]{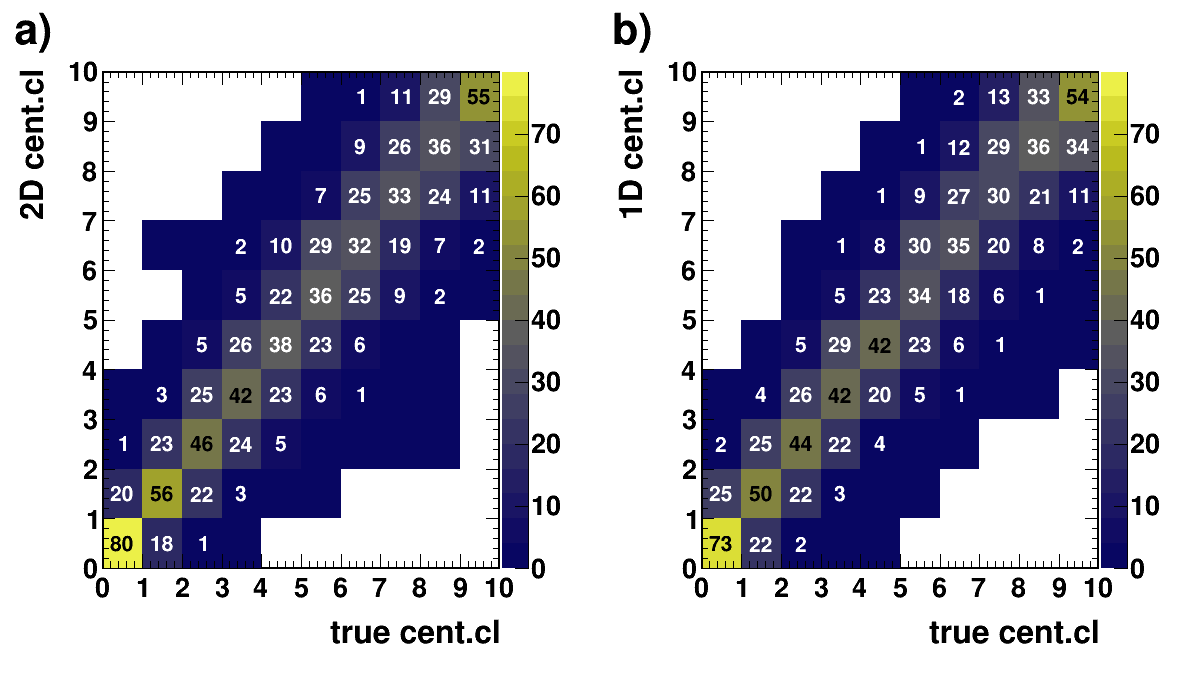}
\caption{Confusion matrix comparing the proposed two-dimensional approach (left) and the one-dimensional method (right) for DCM‑QGSM‑SMM simulated data. The element $M_{ij}$ shows the fraction of events with true centrality class $i$ (based on the impact parameter distribution) that are assigned to reconstructed class $j$.}
\label{fig:CovMatr}
\end{figure}
The centrality dependence of the mean impact parameter is shown in Fig.~\ref{fig:CompMeanImp}.
The error bars in Fig.~\ref{fig:CompMeanImp} represent the standard deviation of the reconstructed impact parameter distribution for the corresponding centrality class, rather than the statistical uncertainty of the mean. Filled rectangles represent the results of the two-dimensional fit for model data from DCM-QGSM-SMM (left) and UrQMD-AMC (right). Open rectangles show the results for the fit of models data with simulation of the trigger efficiency. Results from the one-dimensional fit of $N_{\text{hit}}$ using the method of Ref.~\cite{rogly2018} are indicated by circles. The mean impact parameter distributions for centrality classes defined according to the geometric cross-section from the model (true centrality) are shown with red triangles. The lower panel presents the ratio of the reconstructed mean impact parameter to that corresponding to the true centrality. The one-dimensional fit method proposed in \cite{das2018, rogly2018} relies on the assumption that the observable distribution is normalized to the total event count a condition that cannot be satisfied in actual experimental conditions. Therefore, it is only applicable to minimal‑bias distributions or pure simulation data. 
This explains the systematic deviation seen in Fig.~\ref{fig:CompMeanImp} for the one‑dimensional fit of the trigger‑simulated data, while proposed two‑dimensional approach yields only a small residual difference.
Figure \ref{fig:CovMatr} presents the confusion matrix comparing the performance of the proposed two‑dimensional approach and the method described in Ref.~ \cite{das2018, rogly2018}. Both methods were employed to determine centrality classes, which were then compared with the true classes derived from the simulated impact parameter.
Analysis of the confusion matrix reveals that the proposed two‑dimensional approach achieves higher classification accuracy in the most central collision region compared to the one‑dimensional method. This is reflected by a stronger concentration of events along the main diagonal of the matrix for central bins (e.g., 0–10\%, 10–20\%,20–30\%), indicating better agreement between predicted and true centrality.
The systematic uncertainty associated with the choice of the event generator was estimated by comparing results from DCM-QGSM-SMM and UrQMD-AMC. The differences in the reconstructed mean impact parameter $\langle b \rangle$ between the two models are below 3\% for all centrality classes, indicating that the method is robust against model-dependent effects. Future studies will investigate the influence of different clustering algorithms—such as $k$-means, hierarchical clustering, or graph‑based methods—on the quality of event classification by centrality. Optimizing the clustering procedure may further improve class separation, minimize bin migration, and enhance the overall reliability of centrality determination in the experiment.

\section{Conclusion}

In this work, a two-dimensional Gamma distribution is proposed to characterize the fluctuations of two correlated observables. This distribution accurately describes the two-dimensional distribution of the number of hits in the tracking system and the calorimeter energy deposition for both central and peripheral collisions. Since the track-hit multiplicity $N_{\text{hits}}$ is employed as the second observable, additional selection strategies can be implemented to suppress auto-correlation effects when studying fluctuations of identified particles. These include imposing a pseudorapidity gap ($\eta$-gap) between the track-detection region and the analysis acceptance, as well as excluding protons from the $N_{\text{hits}}$ count in measurements focused on proton-number fluctuations. An alternative extension of the method could involve replacing the number of track hits with the signal from the forward hodoscope as the second observable, which would inherently avoid auto-correlation effects.

The developed distribution forms the basis for an advanced Bayesian method for centrality determination. The performance of this method was validated using simulated data from the DCM-QGSM-SMM and UrQMD model for Xe+CsI collisions at a beam energy of 3.8 $A$ GeV. The reconstructed impact parameter distributions show good agreement with the model predictions across the full centrality range. The centrality dependence of the mean impact parameter in reconstructed data agrees with model simulations, demonstrating the high efficiency of the approach. The proposed approach enables the trigger efficiency of real experimental data to be taken into account in the centrality determination.

Furthermore, multiplicity fluctuations were investigated within a two-parameter Glauber model framework. The analysis reveals that these fluctuations consist of two components: one originating from fluctuations in the number of participant nucleons and binary nucleon-nucleon collisions, and another from fluctuations in the number of particles produced per ancestor. These results enabled the development of a parameterization for fitting that accounts for differences between experimental and simulated observables, consequently reducing the number of free parameters. The proposed methodology is planned for implementation in the BM@N experiment, with future work aimed at extending the suite of observables for centrality determination.

\section{Appendix. Parametrization of observable fluctuations}
Let us estimate the fluctuations of the multiplicity of produced particles within the framework of the two-parameter Glauber model. The multiplicity of produced particles is defined as $N_{ch} =\sum _{i=1}^{N_{a} }n_{i}  $, where $n_{i}-$ is the multiplicity of produced particles per one ancestor, and the number of ancestors is defined as $N_{\text{anc}} = fN_{\text{part}} + (1-f) N_{\text{coll}}$. Thus, in the general case, the multiplicity depends on two random variables - $n$, $N_{\text{anc}}$, then according to the law of total variance
\begin{equation}
\begin{gathered} 
\text{Var}(N_{\text{ch}}) = E[\text{Var}(N_{\text{ch}}|n, N_{\text{anc}})] + \\ E[\text{Var}(E[N_{\text{ch}}|n, N_{\text{anc}}]|N_{\text{anc}})]+ \text{Var}(E[N_{\text{ch}}|n, N_{\text{anc}}])
\end{gathered} 
\end{equation}
where $\text{Var}(N_{\text{ch}}|n, N_{\text{anc}}) = 0$, since $N_{\text{ch}}=Const$ for fixed values of $n$, $N_{\text{anc}}$. The mean number of produced particles for fixed $n$, $N_{\text{anc}}$ is $E[N_{\text{ch}}|n, N_{\text{anc}}] = n N_{\text{anc}}$, then $\text{Var}(E[N_{\text{ch}}|n, N_{\text{anc}}]) =N_{\text{anc}}\sigma_n^2$, where $\sigma_n^2$ is the variance of the multiplicity of produced particles per one ancestor, and finally $E[\text{Var}( N_{\text{ch}}|n, N_{\text{anc}} )] = \langle N_{\text{anc}}\rangle \sigma_n^2$. Now let's define the following expression $E[N_{\text{ch}} | N_{\text{anc}}]= \langle n \rangle N_{\text{anc}}$, where $\langle n \rangle$ is the average multiplicity of produced particles per one ancestor, using it we can compute $\text{Var}(E[N_{\text{ch}}|N_{\text{anc}}]) = \langle n \rangle^{2}\text{Var}( N_{\text{anc}}) $. Finally, we get
\begin{equation}
\text{Var}(N_{\text{ch}}) = \langle N_{\text{anc}} \rangle \sigma_n^2 + \langle n \rangle^2 \text{Var}(N_{\text{anc}})
\end{equation}
Thus, the fluctuations of the multiplicity of produced particles can be divided into two components, the first -- volume fluctuations, due to fluctuations in the number of participants and binary nucleon-nucleon collisions, the second - fluctuations in particle production. The fluctuations in the number of ancestors are determined as
\begin{equation}
\begin{gathered} 
\text{Var}(N_{\text{anc}}) = f^2 \text{Var}(N_{\text{part}}) + (1-f)^2 \text{Var}(N_{\text{coll}}) + \\
2f(1 - f) \text{Cov}(N_{\text{part}}, N_{\text{coll}})
\end{gathered} 
\end{equation}

Suppose we have a certain model of heavy ion collisions in which the number of ancestors $N_{\text{anc}}\simeq N_{\text{anc}}^{\text{MC}}$ is modeled with good accuracy, but the multiplicity of produced particles in the model differs from the experimental one due to minor differences in interaction cross-sections, tracking system efficiency, and temperature effects in the detector (etc.). Then the model will have a distribution $n^{\text{MC}}$, with corresponding mean value and variance $\langle n^{\text{MC}} \rangle$, $\sigma_{n^{\text{MC}}}^2$. Now, based on these assumptions, let's find the relationship between the means and variance of the multiplicity in the experiment and in the model. Let $n = \alpha_{N} n^{\text{MC}}$, then $\langle N_{ch} \rangle = \alpha_{N} \langle N_{ch}^{\text{MC}} \rangle$, let's write the variance of the charged particle multiplicity as follows $\text{Var}(N_{\text{ch}}) =\langle N_{\text{anc}}^{\text{MC}} \rangle \sigma_{n}^{2} +\alpha_{N}^{2}( \langle n^{\text{MC}} \rangle^2 \text{Var}(N_{\text{anc}}^{\text{MC}})+\langle N_{\text{anc}}^{\text{MC}} \rangle \sigma_{n^{\text{MC}}}^{2}-\langle N_{\text{anc}}^{\text{MC}} \rangle \sigma_{n^{\text{MC}}}^{2})$, from the obtained expression we isolate the components responsible for the variance and mean value of the multiplicity in the model $\text{Var}(N_{\text{ch}}) =\langle N_{\text{anc}}^{\text{MC}} \rangle (\sigma_{n}^{2} -\sigma_{n^{\text{MC}}}^{2}\alpha_{N}^{2})+\alpha_{N}^{2}(\text{Var}(N_{\text{ch}}^{\text{MC}}) )=\langle N_{\text{ch}}^{\text{MC}} \rangle (\sigma_{n}^{2} -\sigma_{n^{\text{MC}}}^{2}\alpha_{N}^{2})/\langle n^{\text{MC}} \rangle+\alpha_{N}^{2}(\text{Var}(N_{\text{ch}}^{\text{MC}}) )$, the expression $(\sigma_{n}^{2} -\sigma_{n^{\text{MC}}}^{2}\alpha_{N}^{2})/\langle n^{\text{MC}} \rangle$ is denoted as $\beta_N$, thus we obtain that:
\begin{equation}
\text{Var}(N_{\text{ch}}) = \langle N_{\text{ch}}^{\text{MC}} \rangle \alpha_{N}\beta_N+\alpha_{N}^{2}(\text{Var}(N_{\text{ch}}^{\text{MC}}) )
\end{equation}

Similar reasoning can be applied to the spectator energy, assuming that the number of spectator nucleons in the model well describes the experiment $N_{\text{sp}}=N_{\text{sp}}^{\text{MC}}$, and the energy of one spectator nucleon $e_{\text{sp}}$ scattered in the calorimeter differs from the experimental values only in the mean value and variance, then it is fair to assume that:
\begin{equation}
\text{Var}(E_{\text{sp}}) =  \langle E_{\text{sp}}^{\text{MC}} \rangle \alpha_{E}\beta_E+\alpha_{E}^{2}(\text{Var}(E_{\text{sp}}^{\text{MC}}) )
\end{equation}

\bibliographystyle{JHEP}
\bibliography{bibArt.bib}

\end{document}